\documentclass{jpp}
\usepackage{subeqn}
\usepackage{epsfig}

\let\realverbatim\verbatim
\let\realendverbatim\endverbatim
\renewenvironment{verbatim}
  {\par\addvspace{6pt plus 1pt minus 2pt}\realverbatim}
  {\realendverbatim\addvspace{6pt plus 1pt minus 2pt}}

\ifprodtf \else
  \checkfont{eurm10}
  \iffontfound
    \IfFileExists{upmath.sty}
      {\typeout{^^JFound AMS Euler Roman fonts on the system,
                   using the 'upmath' package.^^J}%
       \usepackage{upmath}}
      {\typeout{^^JFound AMS Euler Roman fonts on the system, but you
                   don't seem to have the}%
       \typeout{'upmath' package installed. JPP.cls can take advantage
                 of these fonts,^^Jif you use 'upmath' package.^^J}%
       \providecommand\umu{\umu}%
      }
  \else
    \providecommand\umu{\mu}%
  \fi
\fi

\ifprodtf \else
  \checkfont{msam10}
  \iffontfound
    \IfFileExists{amssymb.sty}
      {\typeout{^^JFound AMS Symbol fonts on the system, using the
                'amssymb' package.^^J}%
       \usepackage{amssymb}%
         
       \let\ge=\geqslant  \let\geq=\geqslant
      }{}
  \else
  \fi
\fi

\ifprodtf \else
  \IfFileExists{amsbsy.sty}
    {\typeout{^^JFound the 'amsbsy' package on the system, using it.^^J}%
     \usepackage{amsbsy}}
    {}
\fi

\newcommand{\be}{\begin{equation}}
\newcommand{\ee}{\end{equation}}

\newcommand{\eqa}{\begin{eqnarray}}
\newcommand{\eqe}{\end{eqnarray*}}
\newcommand{\eqnu}{\begin{eqnarray}}
\newcommand{\eqne}{\end{eqnarray}}

\newcommand{\eeq}{\end{eqnarray}}

\newdefinition{definition}[theorem]{Definition}

\title[Journal of Plasma Physics]
{Dispersive magnetized  waves in the solar wind plasma}

\author[D. Shaikh]
{B. \ls D\ls A\ls S\ls G\ls U\ls P\ls T\ls A\ls$^2$
\thanks{\tt Email:brahmananda.dasgupta@uah.edu}, \ls
D\ls A\ls S\ls T\ls G\ls E\ls E\ls R \ns S\ls H\ls A\ls I\ls K\ls H\ls$^{1,2}$
\thanks{\tt Email:dastgeer.shaikh@uah.edu} \and \ls P. \ls K. \ls S\ls H\ls U\ls K\ls L\ls A\ls$^3$}
\affiliation{$^1$Department of Physics and \\
$^2$Center for Space Physics and Aeronomic Research (CSPAR),\\
University of Alabama at Huntsville, Huntsville, AL 35805. USA.\\
$^3$Institut f\"ur Theoretische Physik IV, Fakult\"at f\"ur Physik und
Astronomie, Ruhr-Universit\"at Bochum, D-44780 Bochum, Germany}
\date{April 4 2010}
\pubyear{2009} 
\volume{00}
\part{0}
\pagerange{\pageref{firstpage}--\pageref{lastpage}}
\doi{S0963548301004989}

\begin{document}

\label{firstpage}
\maketitle

\begin{abstract}
We derive a generalized linear dispersion relation of waves in a
strongly magnetized, compressible, homogeneous and isotropic
quasineutral plasma. Starting from a two fluid model, describing
distinguishable electron and ion fluids, we obtain a six order linear
dispersion relation of magnetized waves that contains effects due to
electron and ion inertia, finite plasma beta and angular dependence of
phase speed. We investigate propagation characteristics of these
magnetized waves in a regime where scale lengths are comparable with
electron and ion inertial length scales. This regime corresponds
essentially to the solar wind plasma where length scales, comparable
with ion cyclotron frequency, lead to dispersive effects. These scales
in conjunction with linear waves present a great deal of challenges in
understanding the high frequency, small scale dynamics of turbulent
fluctuations in the solar wind plasma.
\end{abstract}


\section{Introduction}
Solar wind plasma is an admixture of waves, structures, and turbulent
fluctuations that comprises multitude of length and time scales. Owing
to a great deal of disparity in the time and spatial scales, solar
wind plasma exhibits rich and complex dynamical evolution. Since the
solar wind plasma is strongly magnetized, the presence of waves and
their consequent interactions with fluctuations complicate our
understanding of many aspects.  For instance, the solar wind
fluctuations yield a composite spectrum [\cite{Goldstein95}] as
described in the schematic of figure 1. This spectrum describes the
power spectral density (PSD) as a function of frequency and can be
divided into five distinct regions. The frequencies smaller than
$10^5$ Hz, namely region I, lead to a PSD that has a spectral slope of
Â­1 [\cite{Goldstein95,MatthaeusGoldstein}]. The region II extends
from $10^5$ Hz to or less than ion/proton gyrofrequency where the
spectral slope exhibits an index of Â­3/2 or Â­5/3. The latter, a
somewhat controversial issue, is characterized essentially by fully
developed turbulence.  The region II connects to the region III by
means of a spectral break at length­scales corresponding to ion
inertial length­scales and frequencies corresponding to ion
gyrofrequencies. The onset of the spectral break is disputed. This
regime is often referred to as dissipative regime which exhibits a PSD
with a much broader spectral slope that varies between ­2 and ­5
[\cite{smith90,Goldstein94,leamon98}]. Notably, the dynamics of the
length­scales in this region cannot be described by the usual
compressible or incompressible MHD models that possess characteristic
frequencies smaller than ion gyro frequencies. A two fluid MHD model
needs to be invoked to examine the dynamics of these high frequency
and fast time scale processes.  What is notable in these spectra is
the waves in different regimes that interact with the turbulent
fluctuations and influence the cascade dynamics
[\cite{Shukla78,Shaikhprl09,Shukla85}]. Despite their prominent
existence in different regimes, less is understood about the evolution
of these waves and their dynamical role in governing the turbulent
spectrum. For example, it is argued that the spectral break results
from the damping of ion cyclotron waves. The latter is contrasted by
Shaikh \& Zank [\cite{dastgeer2}] who report that Hall forces could
play a critical role in producing the spectral break. Similarly, the
excitation and interaction of whistler waves in high frequency
turbulence has been disputed recently
[\cite{Biskamp,dastgeer3,dastgeer4,dastgeer5}].  Motivated by these
issues, we in this paper investigate linear waves in strongly
magnetized plasma that are relevant to the understanding of the solar
wind turbulence spectra depicted in Fig 1. Our main objective here is
to develop a comprehensive understanding of the propagating linear
waves in strongly magnetized plasma especially in the regimes III, IV
and V of Fig 1. We further concentrate on a high beta plasma, because
the solar wind plasma fluctuations are characterized typically by a
high beta plasma beyond 1 AU (astronomical unit).

\begin{figure}[t]
\includegraphics[width=10cm]{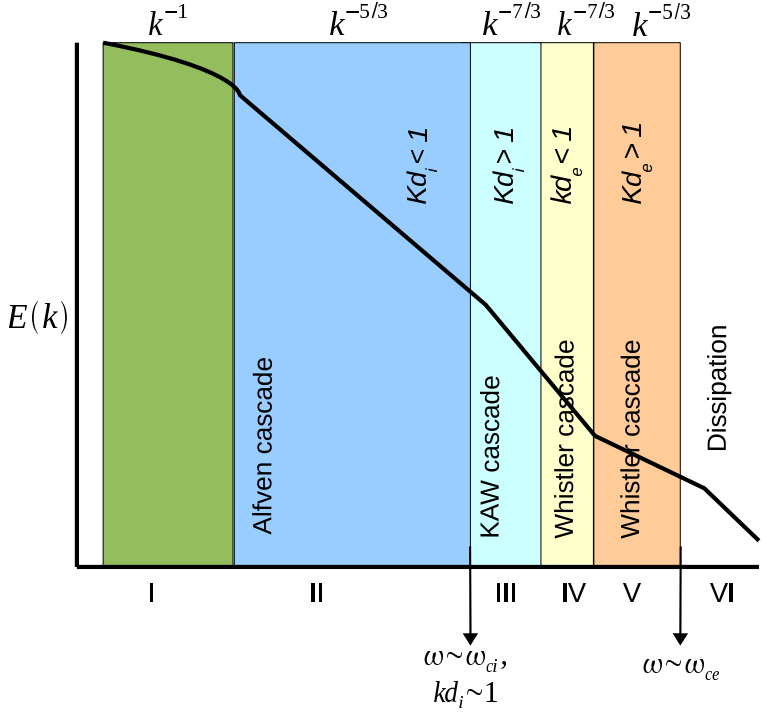}
\caption{Schematic of the solar wind composite spectrum as a function
  of frequency (wavenumber). Magnetized waves play a critical role in
  the transition from region II (MHD regime) to region III (kinetic or
  Hall MHD regime). The onset of regions III and IV is associated with
  high frequency whistler waves where electron and ion motions are
  decoupled.  Regions IV and V are identified as whistler wave cascade
  regimes.}
\label{fig1}
\end{figure}

In section 2, we describe our two fluid model that contains all
possible modes in the magnetized solar wind plasma. A generalized
linear dispersion relation is derived. In section 3, we describe
various possible roots, corresponding to modes.  This section also
describes the solution of the generalized dispersion relation and
focuses on waves especially associated with a high beta (pressure and
magnetic energy ratio) plasma.  Section 4 deals with the effect of
plasma beta on the propagation characteristics of dispersive waves.
Finally, a summary is presented in section 5.

\section{Linear Dispersion Relation for two fluid plasma with finite electron mass}
We start with a two-fluid model (ion with suffix $i$ and electron
with suffix $e$). Using usual notations and with finite electron
mass, i.e., $m_e\neq 0$, the ion and electron continuity
equations, momentum equations, energy equations together with the
Maxwell's equation (neglecting displacement current) are written
as (in CGS Gaussian unit)

\begin{equation}\label{mconie}\frac{\partial n_i}{\partial t}
+\nabla\cdot(n_i {\bf u_i})=0, \quad \frac{\partial n_e}{\partial
t} +\nabla\cdot(n_e {\bf u_e})=0,
\end{equation}

\begin{eqnarray}
m_in_i\left(\frac{\partial {\bf u_i}} {\partial t}+{\bf
u_i}\cdot\nabla{\bf u_i}\right)&=& -\nabla p_i +n_ie\left({\bf
E}+\frac{1}{c}{\bf u_i}\times{\bf B}\right),\label{momconi1}\\
m_en_e\left(\frac{\partial {\bf u_e}} {\partial t}+{\bf
u_e}\cdot\nabla{\bf u_e}\right)&=&-\nabla p_e -n_ee\left({\bf
E}+\frac{1}{c}{\bf u_e}\times{\bf B}\right),\label{momcone1}
\end{eqnarray}

\begin{equation}\label{presie1}\frac{\partial p_i}{\partial t}
+{\bf u_i}\cdot\nabla p_i+\gamma_i p_i\nabla\cdot {\bf u_i} =0,
\quad\frac{\partial p_e}{\partial t} +{\bf u_e}\cdot\nabla
p_e+\gamma_e p_e\nabla\cdot {\bf u_e} =0,
\end{equation}\vspace{-0.5cm}
\begin{eqnarray}
\nabla\times {\bf B}&=& \frac{4\pi e}{c}(n_i{\bf u_i}-n_e{\bf
u_e}), \label{ampere1}\\
\frac{\partial {\bf B}}{\partial t}&=& -c\nabla\times{\bf E},
\label{farad1}\\  \nabla\cdot {\bf B}&=& 0,\label{divb}
\end{eqnarray}
where $n_i, n_e$ are the ion and electron number densities, $m_i,
m_e$ are the ion and electron mass, ${\bf u_i}, {\bf u_e}$ are the
ion and electron velocity, $p_i, p_e$ are the ion and electron
pressure,$\gamma_i, \gamma_e $ are ratio of the specific heats for
ions and electrons,respectively, and {\bf E} is the electric field
and {\bf B} is the magnetic field. We use the linearized forms of
the above equations with the following perturbation scheme for any
variable $\psi_\alpha =\psi_{\alpha0}+\tilde{\psi}_\alpha,
~~(\alpha =i,e$):
\begin{equation}\label{perturb1}
n_\alpha=n_{\alpha0}+\tilde{n}_\alpha;\quad p_\alpha=p_{\alpha0}
+\tilde{p}_\alpha; \quad {\bf u} = {\bf \tilde{u}}_\alpha;\quad
{\bf B}={\bf B_0}+{\bf{\tilde{B}}}~~~{\bf E}={\bf \tilde{E}};
\end{equation}
with $n_{\alpha0}, p_{\alpha0}, B_0 $ are constant and uniform in
space. We assume the space-time dependence of a perturbed variable
$\psi(x, y, z, t)$ as, $\tilde{\psi}\sim \exp[i({\bf k}\cdot {\bf
r}-\omega t)]$. Further, we introduce the plasma displacement
vector $\xi_\alpha$ as, ${\bf \tilde{u}}_\alpha =
\partial {\bf \xi_\alpha}/\partial t  = -i\omega\xi_\alpha$, with
$\xi_\alpha$ as a vector. From the linearized equations for ion
and electron continuity equations, and the energy equations, one
gets, (replacing~ $\partial/\partial t \rightarrow -i\omega; \quad
\nabla \rightarrow i{\bf k}$) the following:

\begin{equation}\label{mconpresie2}
\tilde{n}_\alpha =-in_{\alpha0}({\bf k}\cdot\xi_\alpha), \quad
\tilde{p}_\alpha =-i\gamma_\alpha p_{\alpha0}({\bf k}
 \cdot\xi_\alpha),
\end{equation}

The linearized form of (\ref{ampere1}) after   combining  with
linearized form of (\ref{farad1}), can be written as, (with
$n_{i0}=n_{e0}=n_0 $),

\begin{equation}\label{ampere2}
{\bf k}\times{\bf k}\times{\bf \tilde{E}}= -\omega^2\frac{4\pi
n_0e}{c^2}({\bf \xi_i}- {\bf \xi_e})
\end{equation}

Finally, we write the linearized forms of momentum conservation
equations as:

\begin{eqnarray}
-\omega^2 m_in_{0}{\bf \xi_i}&=& -\gamma_i p_{i0}{\bf k}({\bf k}
\cdot{\bf \xi_i}) +n_0e\left({\bf \tilde{E}}-\frac{i\omega}{c}
~{\bf \xi_i}\times{\bf B_0}\right),\label{momconi2}\\
-\omega^2 m_en_{0}{\bf \xi_i}&=&-\gamma_e p_{e0}{\bf k}({\bf k}
\cdot{\bf \xi_e})+n_0e d_e^2~{\bf k}\times{\bf k}\times{\bf
\tilde{E}} -n_{0}e\left({\bf \tilde{E}}-\frac{i\omega}{c}~{\bf
\xi_e}\times{\bf B_0}\right),\label{momcone2}\nonumber\\
\end{eqnarray}
where we used $n_{i0}= n_{e0} =n_0$ and replaced ${\bf\xi_e} $ the
displacement vector for electron by the relation obtained from
(\ref{ampere2}), i.e.,$${\bf \xi_e}= {\bf \xi_i}+\frac{c^2}{4\pi
n_0e\omega^2}~{\bf k}\times{\bf k}\times{\bf \tilde{E}};$$ and
introduced the electron inertial length $d_e = c/\omega_{pe}$,~
$\omega_{pe}$ being the electron plasma frequency. From  eqn.
(\ref{ampere2})taking a dot-product with {\bf k}, and also using
the continuity equation (\ref{mconpresie2}) we get
\begin{equation}\label{quasin}
{\bf k} \cdot{\bf \xi_i} = {\bf k} \cdot{\bf \xi_e},\qquad
\tilde{n}_e =\tilde{n}_i,
\end{equation}

where the second equation denotes the quasineutrality condition.
Equations (\ref{momconi2}) and (\ref{momcone2}) are our starting
equations for deriving the dispersion relation for a two fluid
plasma. We followed the procedures adopted by Ishida et al., to
get the dispersion relation and we omit the derivation in this
work for brevity. We proceed by adding both the equations,
eliminating the perturbed electric field ${\bf \tilde{E}}$ in
terms of ${\bf\xi_i}$ using (\ref{momconi2}) and finally taking
the dot product of the resulting equation for ${\bf\xi_i}$ with
three independent vectors, ${\bf k},~ {\bf{\hat{b}}} $ and ${\bf
k}\times {\bf{\hat{b}}}$, where ${\bf{\hat{b}}}$ is the unit
vector along the unperturbed magnetic field ${\bf B_0}$. The final
generalized dispersion relation reads as

\begin{eqnarray}\label{dispersie1}
&&\left[\omega^2(1+k^2 d_e^2)-k^2V_A^2\cos^2\theta
\right]\left[\omega^4-k^2(V_A^2+C_S^2)\omega^2
+k^2V_A^2k^2C_S^2\cos^2\theta\right.\nonumber\\
&&\left.+\omega^2k^2 d_e^2(\omega^2-k^2C_S^2)\right]=
\omega^2V_A^2k^2d_i^2 k^2\cos^2\theta (\omega^2-k^2C_S^2);
\end{eqnarray}
where $d_i= c/\omega_{pi}$, $\omega_{pi}$ being the ion plasma
frequency, and $\theta$ is the angle between the wave vector {\bf
k} and unperturbed magnetic field ${\bf B_0}$, and $V_A, ~ C_S$
are the Alfv\'{e}n speed and sound speed, respectively. This
dispersion relation (\ref{dispersie1}) incorporates both the ion
and electron inertial length scales, $d_i$ and $d_e$,
respectively. Moreover, equation (\ref{dispersie1}) is essentially
the same as the dispersion relations obtained by earlier authors
in Refs [\cite{Stringer,Swanson,Damiano}]. It is easy to see that
for electron mass $m_e\rightarrow 0$, which is equivalent to
putting $d_e=0 $, eqn. (\ref{dispersie1}) reduces to
 \begin{eqnarray}\label{dispersishi}
&&\left[\omega^2-k^2V_A^2\cos^2\theta
\right]\left[\omega^4-k^2(V_A^2+C_S^2)\omega^2
+k^2V_A^2k^2C_S^2\cos^2\theta\right]\nonumber\\ && =
\omega^2V_A^2k^2d_i^2 k^2\cos^2\theta (\omega^2-k^2C_S^2);
\end{eqnarray}
which is the usual Hall MHD dispersion relation obtained by Ishida
et al. [\cite{Ishida}]. For both $d_i=0, ~d_e=0 $ eqn.
(\ref{dispersie1}) reduces to the dispersion relation for an ideal
MHD, showing the existence of shear Alfv\'{e}n wave, fast and slow
magnetosonic
waves. \\

From the dispersion relation (\ref{dispersishi}) Ishida et al
[\cite{Ishida}] have shown some interesting effects of the presence
of ion inertial length. It is shown that for the limit $kd_i > 1$,
the incompressible MHD Alfv\'{e}n wave becomes compressible and
the MHD compressible slow wave becomes incompressible. While we
postpone similar studies for the inclusion of electron inertial
length for our future work, we shall study the effects of plasma
$\beta$ on the linear waves in solar wind from the dispersion
relation (\ref{dispersie1}). For that we normalize the frequency
$\omega$ by some characteristic frequency $\omega_0 $,so that the
dimensionless frequency $\bar{\omega}=\omega/\omega_0 \equiv
\omega$. Let $U_0$ be some characteristic speed. Thus
$L_0=U_0/\omega_0 $ will be some characteristic length, Choosing
$\omega_0 \equiv (kV_A)$, and note that
$$\frac{C_S^2}{V_A^2}= \frac{\gamma p_0/\rho_0}{B_0^2/4\pi\rho_0}
= \frac{4\pi\gamma p_0}{B_0^2}\approx \frac{8\pi p_0}{B_0^2}=
\beta, =\frac{M_A^2}{M_S^2}$$ where $M_A$ and $M_S$ are
Alfv\'{e}nic and sonic Mach numbers.

We next normalize $\omega$ by $kV_A$ and $\omega/k = v_p$, the
phase velocity, and $\omega/kV_A$ is the normalized phase
velocity. The linear phase velocity relation can then be expressed
as

\begin{eqnarray}\label{dispdimless}
&&\left[v_p^2(1+(k d_e)^2)-\cos^2\theta \right]=
\left[v_p^4-(1+\beta)v_p^2 +\beta\cos^2\theta
+v_p^2(kd_e)^2(v_p^2-\beta)\right]\nonumber\\
&&=v_p^2(kd_i)^2 \cos^2\theta(v_p^2-\beta)
\end{eqnarray}

So the final form of the dispersion relation given by
eqn.(\ref{dispdimless}) is essentially an equation showing the
angular dependence of normalized phase velocity to plasma $\beta$
and the ion electron inertial length scales.

\section{Dynamics of linear dispersive waves}
We have developed a matlab code to solve the generalized
dispersion relation Eq. (\ref{dispdimless}). The dispersion
relation is 6th order in frequency. Hence six roots are expected.
Three roots correspond to forward and the remaining three
represent backward propagating waves. The generalized dispersion
relation, Eq. (\ref{dispdimless}), spans a wider parameter regime
and exhibits a variety of waves for various angle, beta,
length-scales comparable with the electron and ion inertial skin
depths. In this paper, we nonetheless restrict ourselves to the
parameter regime that is a representive of the solar wind plasma
corresponding to the length scales that are associated with the
ion cyclotron frequency.

\begin{figure}[t]
\begin{center}
\epsfig{file=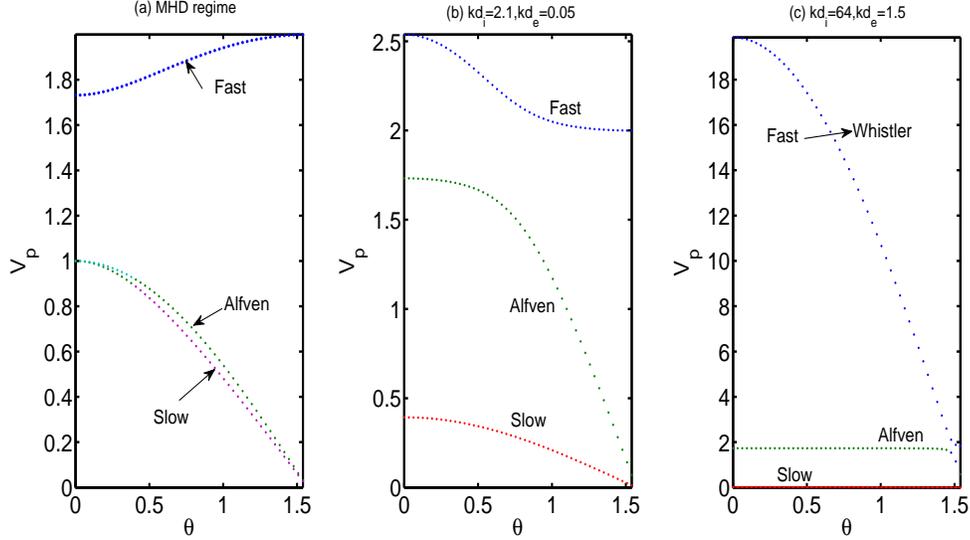, width=14.cm, height=8.cm}
\end{center}
\caption{(a) Linear waves in MHD regime. Forward propagating fast,
slow, and Alfv\'enic
  modes are shown. The top curve describes Alfv\'enic mode, whereas
  the remaining two curves are overlapped and represent fast and slow
  modes. Our results are consistent with Ref [\cite{amitava}]. (b) High
  frequency waves in $kd_i>1$ and $kd_e<1$ (c) $kd_i>1$ and $kd_e>1$
  regimes. It is clear from (b) and (c) that the propagation
  characteristic is modified dramatically in the short scale
  regimes. The phase speed increase for shorter scales. The parallel
  propagation is faster than the oblique one.}
\label{fig2}
\end{figure}

As a first step, we verify the consistency of our equation by
comparing it with the MHD waves \cite{amitava}.  For this purpose,
we use $d_i=d_e=0$ that reduces the dispersion relation to the
usual compressible MHD relation. We retrieve MHD waves that are
consistent with Ref \cite{amitava}. This is shown in Fig (1a) that
describes a phase velocity variation of the MHD waves as a
function of angle between the propagation wave vector and the mean
magnetic field, i.e. $k_\parallel = k \cos \theta$ where
$k_\parallel = {\bf k} \cdot {\bf B}_0$. We find three forward
propagating modes with positive phase velocity that co-exist with
three backward propagating modes with negative phase velocity. It
is noteworthy that the shear Alfv\'{e}n and slow modes are
partially overlapped. In this regime, linear dynamics is entirely
governed by the magnetosonic waves which is shown by the top curve
in Fig (1a). A pure magnetic perturbation propagating orthogonal
to the constant magnetic field in this regime behaves
electrostatically and tends to move as a magnetosonic mode.

We next investigate the effect of a finite plasma beta on the
propagation of the small scale magnetized waves to explore their
dynamics in regimes III and IV in Fig (1). We choose parameters
$\beta=3, k=0.5, d_e=0.1, d_i=43d_e$ that are characterize the
high frequency modes between regions II and III. Since this
parameter regime is not drastically different than the typical MHD
regime, a non trivial modification in the dispersion of MHD waves
is expected. For instance, the linear dynamics in this regime is
governed predominantly by the higher frequency and short scale
waves, namely kinetic Alfv\'en and magnetosonic waves.  The
density perturbation becomes significant in this regime.  The
dispersion of high frequency kinetic Alfv\'en, shown in Fig (2b),
differs significantly from that in Fig (2a). The phase velocity of
the kinetic Alfv\'en waves is increased, while the obliquely
propagating kinetic Alfv\'en modes continue to remain unaffected.
The shear Alfv\'{e}n/slow modes decouple clearly. The fast modes
travel with higher phase speed, whereas the slow modes propagate
slower (than in Fig 2a). It is further clear from Fig (2b) that
the propagation of the fast/slow modes depends crucially on the
alignment of their wave vector relative the mean magnetic field.
For a highly oblique propagation, these waves hardly move.  Figure
(2c) describes a regime where $kd_i>1$ and $kd_e \ge 1$.  This is
a regime where MHD modes are drastically altered and low frequency
whistler modes start to play critical role. Shown in Fig (2c), the
top curve is the whistler branch that survives whereas the bottom
curves are reminiscent of MHD modes (Alfv\'en and fast/slow). As
seen in Fig (2c), the whistler modes have higher frequency and
phase speeds. They propagate predominantly along the field lines,
whereas oblique whistlers have smaller phase speeds. At an angle
$\theta=\pi/2$, the whistler modes do not propagate. They are
transformed into high frequency electrostatic modes. Figure (2c)
describes waves that possess length scales smaller than the ion
inertial length, but bigger than electron inertial length scale.
We can retrieve the small scale waves by choosing the small
characteristic length scales of these waves relative to the ion
inertial and electron inertial length scales. This is shown in Fig
(3). This figure is further consistent with Fig (2) where the
dynamics of small scale is described by choosing a relatively
small magnitude of $k$ compared to $d_i$ and $d_e$ lengths.


\begin{figure}[t]
\includegraphics[width=12cm]{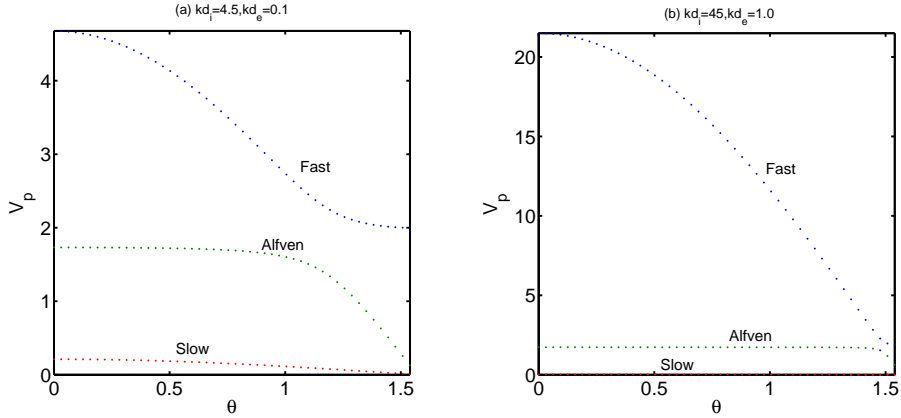}
\caption{Figures (a) and (b) are shown for a large $k$. Note that
the phase velocity of the small scale wave is consistent with Fig
(2).} \label{fig3}
\end{figure}


\begin{figure}[t]
\includegraphics[width=12cm]{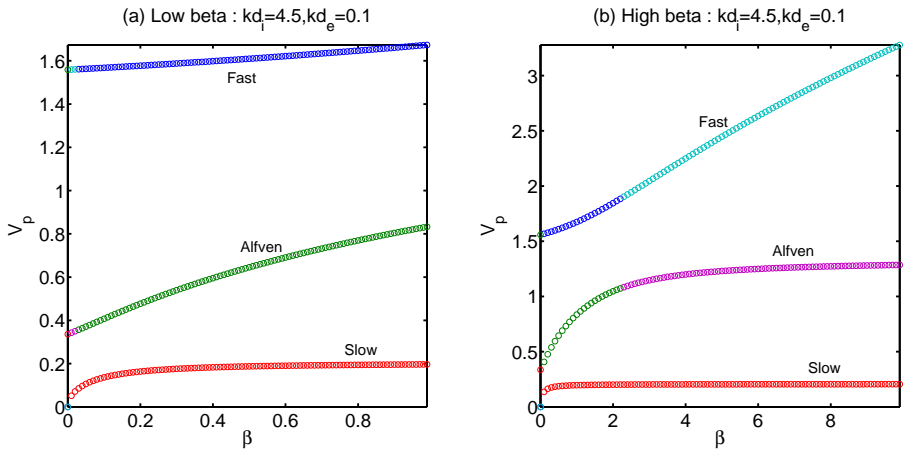}
\caption{Effect of plasma beta is shown for large scale waves.
(a) describes a low plasma beta regime, whereas (b) represents the higher
beta regime.} \label{fig4}
\end{figure}

\section{Effect of plasma beta on the dispersive waves}
We next investigate effect of plasma beta (ratio of plasma pressure
and magnetic energy) on the propagation of dispersive waves in the
solar wind plasma. We keep the angle of propagation constant
($\theta \simeq 45^\circ$) and vary plasma $\beta$ and study the waves
corresponding to the small and large scale modes. A snap shot for the
large scale waves is shown in Fig (4) that depicts effects of low (Fig
4a) and high (Fig 4b) plasma beta on the propagating waves. We find
that the large scale Alfv\'en and slow waves continue to remain
unaffected with regard to the beta varation. The propagation speed of
the fast waves appears to grow linearly with the plasma beta. Hence
there is a dramatic speed enhancement occurs for the fast waves. This
is shown in Fig (4b).

Interestingly, we find that the propagation property of the dispersive
waves corresponding to the small scale modes does not depend
critically on the plasma beta. Almost all of these small scale waves
remain unaffected relative to the beta variation. This is shown in Fig
(5a) for the low beta regime, whereas Fig (5b) describes the effect of
high plasma beta on the small scale dispersive waves.

\begin{figure}[t]
\includegraphics[width=12cm]{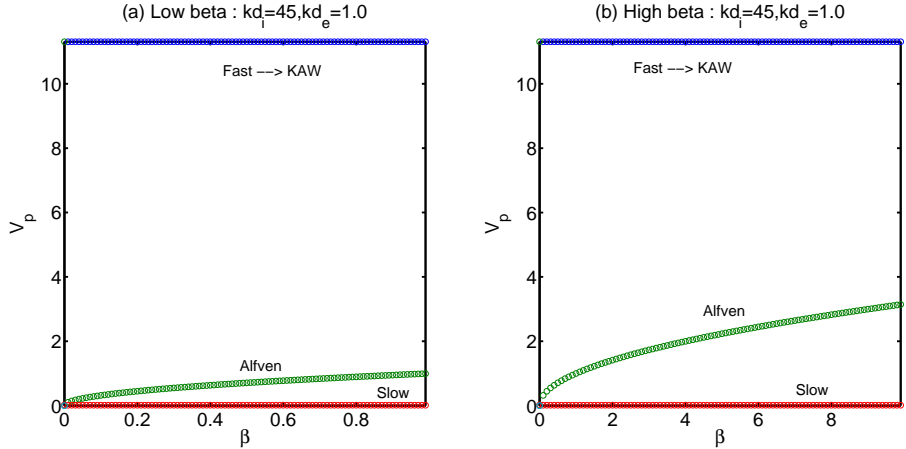}
\caption{Phase speed variation with respect plasma beta for small scale
dispersive waves. (a) and (b) describe respectively low and high
beta regimes.} \label{fig5}
\end{figure}

\section{Summary}
A major outcome of work is the inclusion of electron and ion
inertia in strongly magnetized waves associated with a high beta
plasma. We invoked a two fluid or Hall MHD model to examine the
dynamics of these high frequency and fast time scale processes. We
see that for the length scales $kd_e\geq 1$, where there the
turbulent spectra shows a different power law behavior, the linear
waves are significantly modified. It is clear from our work that
the evolution of the linear waves in the MHD regime is modifed
significantly in the dissipative or dispersive regime where
$kd_e>kd_i>1$. This is the regime where solar wind turbulence is
affected non trivially by the linear collision less waves. We find
from our linear theory that MHD modes, such as Alfv\'en and
fast/slow waves, are inconsequential in the small scale
$kd_e>kd_i>1$ regime. This is because the MHD modes are relatively
large scale and low frequency modes in comparison with the kinetic
Alfv\'en and/or whistler modes that are typically excited in the
$kd_e>kd_i>1$ regimes of Fig (1). Because of the great disparity
in the length and time scales, the waves in the two distinct
regimes (i.e. regimes III, IV and V in Fig 1) do not couple
efficiently. Consequently, the linear dynamics of modes in the
regimes IV and V is governed predominantly by fast and whistler
modes respectively. The latter have a greater phase velocity
relative to the MHD modes. Our results are important particularly
to understand linear waves in the solar wind turbulence in the
regime where small scale fluctuations exhibit dispersive
characteristics
[\cite{Goldstein95,MatthaeusGoldstein,leamon98,smith90,dastgeer2,dastgeer3,dastgeer4,dastgeer5}].

\section{Acknowledgment} This research was partially supported by
the NASA(NNX-08AE41G) NASA(NNG-05GH38) and NSF (ATM-0317509)
grants.


\end{document}